\documentclass[%
 reprint,
 amsmath,amssymb,
 aps,
nofootinbib
]{revtex4-2}

\usepackage{revquantum}
\usepackage{graphicx}
\usepackage{dcolumn}
\usepackage{bm}
\usepackage{upgreek}
\usepackage{xcolor}
\usepackage{enumitem}
\usepackage{placeins}
\usepackage{braket}
\usepackage{siunitx}
\newcommand{\red}[1]{\textcolor{red}{#1}}

\usepackage{dsfont}
\usepackage{comment}

\begin{document}

\title{Optical control of the complex phase of a quantum ground state amplitude}

\author{Adam Kinos}
\email{adam.kinos@fysik.lth.se}
\address{Department of Physics, Lund University, P.O. Box 118, SE-22100 Lund, Sweden
}

\author{Mogens Dalgaard}
\address{Center for Complex Quantum Systems, Department of Physics and Astronomy, Aarhus University, Ny Munkegade 120, DK-8000 Aarhus C, Denmark
}

\author{Klaus M{\o}lmer}
\email{moelmer@phys.au.dk}
\address{Center for Complex Quantum Systems, Department of Physics and Astronomy, Aarhus University, Ny Munkegade 120, DK-8000 Aarhus C, Denmark \\
Aarhus Institute of Advanced Studies, Aarhus University, H{\o}egh-Guldbergs
Gade 6B, DK-8000 Aarhus C, Denmark
}
\date{\today}

\begin{abstract}
We discuss how coherent driving of a two-level quantum system can be used to induce a complex phase on the ground state and we discuss its geometric and dynamic contributions. While the global phase of a wave function has no physical significance, coherent dynamics in a two-level subspace provides relative phases and is an essential building block for more advanced dynamics in larger systems. In this regard, we note that one must be careful with intuitive accounts of the phase dynamics as it depends on the interaction picture applied. To mitigate ambiguities in practical analyses, we suggest to complement the Bloch sphere picture with the path taken by the ground state amplitude in the complex plane, and we show how the two-level pure state dynamics can serve as a starting point for the study of the dynamics explored in three-level lambda systems, four level tripod systems, and open quantum systems. 
\end{abstract}

\maketitle

\section{\label{sec:intro}Introduction}


The complex wave function and the variation of its phase sets quantum mechanics fundamentally apart from classical mechanics. A quantum particle encodes its momentum content in the spatial phase variations of its wave function, and relative phases between the different eigenstate amplitudes directly affect expectation values of, e.g., dipole moments and transition operators. In the context of quantum computing, the phase gate employs the ability to control the phase of a single state component of a qubit as an elementary building block from which other, more elaborate operations can be constructed. As is well known, a global phase factor does not change the physical content of a quantum state; only relative phases do. However, employing ancillary degrees of freedom, such as different paths followed by spin particles, ancillary levels in a single particle, or dynamics conditioned on the state of ancillary qubits, has made it possible to observe inherently global phase factors such as the famous sign change acquired by the complete rotation of a spin $1/2$ particle \cite{Stoll1978,Rauch1975,Wood2020}.


This article is concerned with how the phase of a quantum state can be controlled by it coherent coupling to another state. The work is motivated by quantum information processing with laser driven atoms and ions, and we may readily solve the full quantum dynamics of a laser-driven two-level system analytically in simple cases and predict the accumulated phase based on simple geometric arguments. However, when considering more elaborate control schemes that, e.g., minimize the degree of excitation or incorporate robustness against variation of the physical parameters, the accumulated phase is no longer straightforward to predict.

In 1984, Berry famously decomposed the phase evolution of a quantum state into a dynamic and geometric part for cyclic adiabatic processes \cite{Berry1984}. Later, Aharonov and Anandan generalized this result to arbitrary (non-adiabatic) cyclic processes \cite{Aharonov1987}, and, inspired by work by Pancharatnam \cite{Pancharatnam1956}, this decomposition was also established for non-cyclic processes \cite{Samuel1988,Pati1995,Pati1995a}. The geometric, or Berry, phase is often regarded as particularly robust due to its geometric properties, and with extension to non-Abelian phases it has been proposed, both as a practical and as a foundational element in quantum computing \cite{Zanardi1999, Pachos1999, Recati2002, Eisert2003}. For a recent review, see Ref.~\cite{Sjoqvist2015}. The geometric phase has a deep connection with the mathematical theory of fibre bundles and holonomies \cite{Wu1975}, which has also been pursued in polarization optics \cite{Cisowski2022}. 

While being of foundational interest and application in its own right, we note that the distinction of geometric and dynamic phases is not unique and hence a potential source of confusion in the design of, e.g., phase gates. Two-state dynamics are often visually represented by trajectories on the surface of a Bloch sphere, but the physically relevant, global phase acquired by the quantum state is not uniquely identified by the shape or the area enclosed by these trajectories. 

The present work is motivated by the proposal to implement quantum computing with rare-earth ions in crystal materials \cite{Ohlsson2002, Wesenberg2007, Walther2015, Kinos2021}. In these systems, qubits are encoded in hyperfine ground states, and the inhomogeneously broadened transitions to electronically excited states allow addressing individual qubits by their optical excitation frequency \cite{Kinos2021a, Kinos2022, Kinos2022a, Kinos2021}. In this article we specifically analyze the phase acquired by a ground state amplitude under a cyclic optical excitation process in a two-level system, and then proceed to derive control protocols that achieve the same in more complicated three-level lambda systems and higher-dimensional systems. The blockade mechanism induced by the dipole-dipole interaction of nearby ions permit schemes for two-qubit controlled gates and provides access to universal quantum computing \cite{Roos2004, Kinos2021a}. We recall that multiplying a phase factor on a single ground state amplitude effectively acts as a gate operation on a qubit system formed by two or more ground states. Similarly, if the phase acquired by a ground state in one ion depends on the state of a neighbour ion, it formally constitutes a controlled phase gate. Although our initial motivation originated from rare-earth-ion-doped crystals, our analysis applies to other systems where coherent driving is a convenient way to couple or phase shift qubit amplitudes. Note, in particular the similarity with the Rydberg blockade gate for neutral atoms \cite{Jaksch2000, Urban2009, Gaetan2009}. 


We organize this article as follows. In Sec. \ref{sec:phase}, we present the dynamics of a two-level system subject to coherent driving, and we introduce a plot of the complex ground state amplitude, which, unlike the Bloch sphere, keeps track of the acquired phase. We compare four different driving schemes and highlight the interplay of dynamic and geometric phases. In Sec. \ref{sec:lambda_tripod}, we introduce an additional auxiliary long lived state and derive control protocols that exploit the more rich level structure. Sec. \ref{sec:diss} presents ways to discuss the phase of a quantum state in the presence of dissipation, and Sec. \ref{sec:conc} summarizes our conclusions and presents a brief discussion of their main consequences.

\section{Putting a phase on an atomic ground state wave function}\label{sec:phase}

\begin{figure*}
    \centering
    \includegraphics[width=\textwidth]{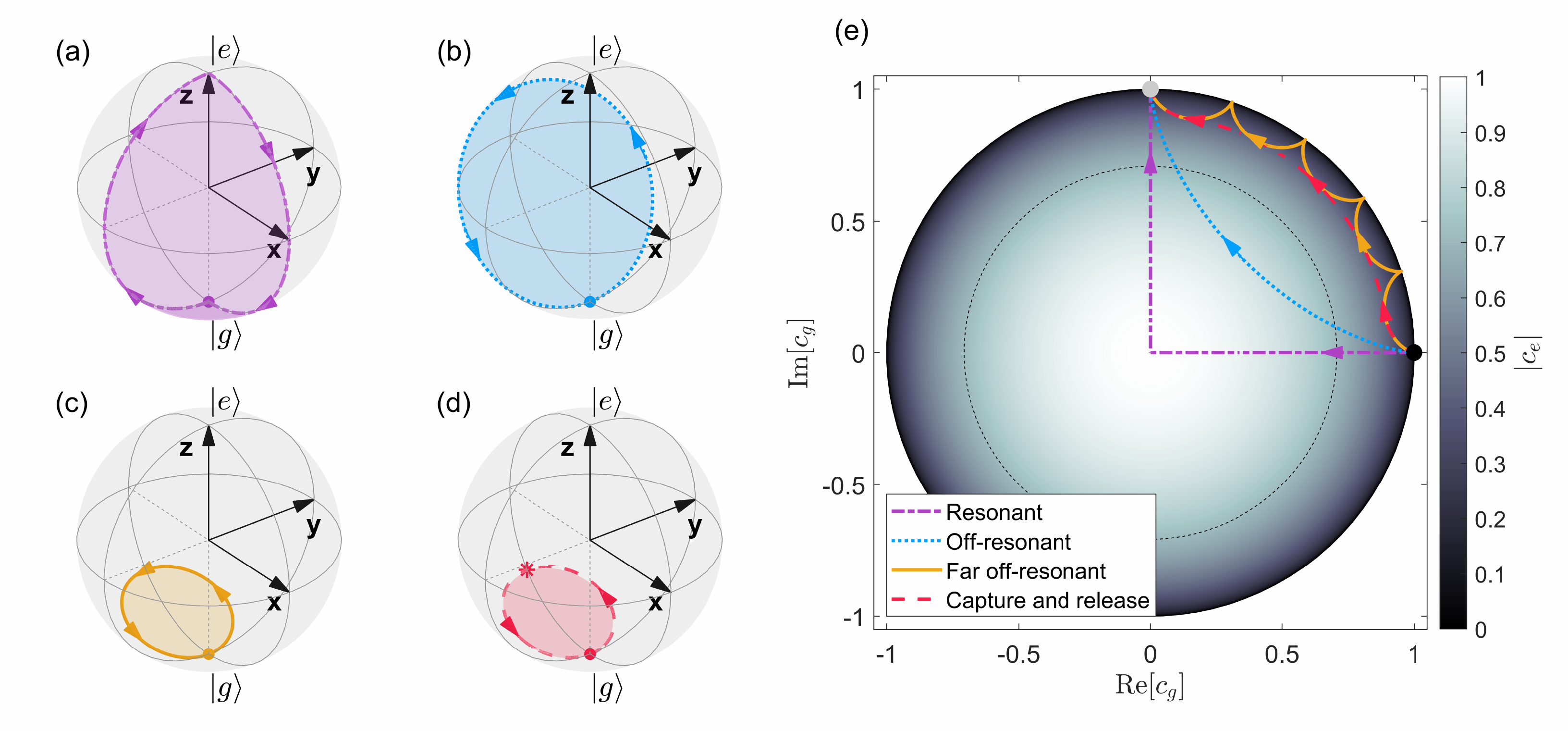}
    \caption{Illustrations of the Bloch vector trajectory for four operations; (a) resonant drive, (b) off-resonant drive, (c) far off-resonant drive, and (d) capture and release of the energy eigenstate, are shown in the interaction picture described by Eq.~(\ref{eq:2level_Hamiltonian}). All operations induce a phase $\theta = \pi/2$ on $\ket{g}$ (the trajectory in panel (c) is encircled 5 times). (e) Depicts the same operations in a complementary $c_g$-plot, which shows the real and the imaginary part of the complex amplitude $c_g$. The ground state population, $|c_g|^2$, is read from the figure as the square of the distance from the center of the disc ($1$ at the disc's edge, $1/2$ at the dashed black circle, and $0$ in the center). The color gradient indicates the amplitude of the excited state coefficient $|c_e|$. The phase $\theta$ is read as the angle between the initial (black dot) and the final (gray dot) state amplitudes. }
    \label{fig:bloch_sphere_illustration}
\end{figure*}

In this section, we consider an atomic two-level system with the ground, $\ket{g}$, and excited state, $\ket{e}$. For convenience, we set the ground state energy to zero and the excited state to $\hbar\omega_A$. Hence, the ground state amplitude does not evolve in the absence of an external drive. In the presence of an external classical drive with frequency $\omega=\omega_A-\Delta$ and Rabi frequency $\Omega e^{i\phi}$, the atomic system's Hamiltonian in the rotating frame (interaction picture with respect to $H_0=\hbar\omega \ket{e}\bra{e}$) is 
\begin{align}
    H = 
    \begin{bmatrix}
    \Delta & \frac{\Omega}{2} e^{i\phi} \\
    \frac{\Omega}{2} e^{-i\phi} & 0
    \end{bmatrix},
    \label{eq:2level_Hamiltonian}
\end{align}
where we have used the rotating wave approximation and set $\hbar = 1$. 

The solution to the Schr\"odinger equation is simple in the specific case of a constant drive, i.e., constant detuning $\Delta$ and Rabi frequency $\Omega e^{i\phi}$, and yields the time-dependent state amplitudes (assuming the initial state $c_g(0)=1$)
\begin{align} \nonumber
    c_e(t) &= -i \frac{\Omega e^{i\phi}}{\Omega_R}
	\sin \bigg(\frac{\Omega_R }{2} t\bigg) e^{-i\frac{\Delta}{2} t}, \\
	c_g(t) &= \Bigg( 
	\cos \bigg( \frac{\Omega_R }{2} t \bigg)
    + i \frac{\Delta}{\Omega_R}
	\sin \bigg( \frac{\Omega_R }{2} t \bigg)
	\Bigg) e^{-i\frac{\Delta}{2} t},
	\label{eq:bloch_dynamics}
\end{align}
where $\Omega_R = \sqrt{\Omega^2 + \Delta^2}$ denotes the generalized Rabi frequency. When $\Omega_R t$ is a multiple of $2\pi$, the excited state population vanishes, and the ground state has unit amplitude with an additional phase factor, which we may control by carefully designing the external drive. We recognize in particular the notable change of sign of the ground state amplitude by a resonant ($\Delta=0$) $2\pi$-pulse due to the sign change of the cosine function. 

A common way to show the evolution of a two-level system is to use the Bloch sphere, as shown in Fig.~\ref{fig:bloch_sphere_illustration}(a-d). The Bloch vector, $\vec{r}$, represents the density matrix elements of the quantum state of a two-level quantum system, where the vertical $z$ coordinate and the longitude azimuthal angle depict the population difference and relative phase between the state amplitudes $c_g(t)$ and $c_e(t)$, respectively. Subject to the coherent driving field, the Bloch vector is exposed to a torque cf., the equation, $d\vec{r}/dt = \vec{W}\times \vec{r}$, where $\vec{W} = [\Omega \cos(\phi), -\Omega \sin(\phi), \Delta]^T$. Note, however, that the Bloch vector provides no information about the phase $\theta$ accumulated by the ground state amplitude $c_g(t)$ during an operation that returns the system to $\ket{g}$. 

Since the otherwise useful Bloch vector does not provide the global phase $\theta$, we supplement the Bloch sphere picture with the entire time dependence of the complex amplitude $c_g(t)$. It is convenient to show this as the evolution of a two-dimensional vector ($\text{Re}[c_g]$, $\text{Im} [c_g]$) in the complex plane, cf., Fig.~\ref{fig:bloch_sphere_illustration}(e). The accumulated phase $\theta$ is the angle between the initial and final amplitudes in the complex plane, while the population in $\ket{g}$ is the square of the distance from the origin. The ground state population, $|c_g|^2$, equals unity on the disc's edge, $1/2$ on the dashed line, and $0$ at the origin. The gradient color scaling visualizes the absolute value of the excited state amplitude, $|c_e| = \sqrt{1-|c_g|^2}$. From the Schr\"odinger equation, we have $|\dot{c}_g| = \frac{\Omega}{2}|c_e|$, which demonstrates that for a given Rabi frequency, $c_g$ changes faster near the center of the disc (lighter regions) and slower near the edge of the disc (darker regions). As we show in Appendix \ref{app:harmonic_oscillator}, the evolution of the complex amplitude $c_g$ has an analogy with the intuitive picture of a long swinging pendulum. 

\subsection{\label{sec:examples}Four examples}


The induced phase $\theta$ on $\ket{g}$ is controllable in many ways, but if, for instance, the excited state is short-lived, it may be preferable to limit the integrated excited state population during the operation. On the other hand, spectator levels in real systems may be off-resonantly excited and give rise to phase errors and population leakage if the Rabi frequency is too high or the detuning to other levels is not high enough. If an operation is conditioned on whether another (control) atom is excited or not, the total duration of the operation should be kept shorter than the lifetime of the control atom and, likewise, unwanted environmental interactions may lead to dephasing over time if the entire control duration is not limited. The state evolution depicted in the $c_g$-plot can partly help assess these concerns. In the following, we introduce and compare four strategies to control the complex phase on $\ket{g}$ and we discuss their potential advantages and disadvantages.

\subsubsection{Two resonant pulses}

Our first example consists of two resonant pulses that bring the system from the ground state $\ket{g}$ to the excited state $\ket{e}$ $(\phi = 0)$ and back again with a relative phase $(\phi = \pi-\theta)$ \cite{Roos2004, Kinos2021a}. The evolution on the Bloch sphere and the $c_g$-plot is shown in Fig. \ref{fig:bloch_sphere_illustration}(a, e) (dash-dotted purple lines). In both plots, it is clear that we transfer all population, and hence accumulate a large integrated population, in the excited state during the operation. By applying a time-dependent Rabi frequency it is possible to pass between the ground and excited state with more advanced composite or tailored pulses from ESR and NMR spectroscopy, such as sechyp or HSH pulses, which are robust against certain imperfections \cite{Roos2004, Tian2011}. The integrated population of the excited state obeys the inequality
\begin{align}
    \int_0^T |c_e(t)|^2 dt \geq \frac{\pi}{\Omega_{\max}},
\end{align}
where $\Omega_{\max}$ denotes the maximum Rabi frequency. Here the equality sign corresponds to piecewise constant pulses with an instantaneous phase change. In the more realistic case of smoothly changing pulses, the integrated population in the excited state becomes larger than this value. 

\subsubsection{Off-resonant driving\label{sec:off_res}}

The second operation, also treated in Refs.~\cite{Suter1988,Wood2020}, drives the system off-resonantly with a fixed ratio $\Delta/\Omega$ such that the state vector traces out a cone in the Bloch sphere. Hence, Eq.~(\ref{eq:bloch_dynamics}) describes the system dynamics, where we depict the corresponding trajectory in Fig.~\ref{fig:bloch_sphere_illustration}(b, e) (dotted blue lines). For constant Rabi frequency, $\Omega$, and detuning, $\Delta$, the state returns after a time $T = 2\pi/\Omega_R$ with a phase $\theta = \pi (1-\frac{\Delta}{\Omega_R})$. Here the integrated excited state population during the process is $\int_0^T |c_e(t)|^2 dt = \pi \Omega^2/\Omega_R^3$. The $c_g$-plot clearly shows that the state traverses a shorter path than the resonant drive and mainly evolves rapidly through the lighter regions of the disk. Unlike the resonant drive, the smooth trajectory indicates that we do not require sudden changes in the driving phase.

\subsubsection{Far off-resonant driving}

Our third operation is similar to the second one, except we explore a much larger detuning, $\Delta/\Omega \gg 1$ while repeating the traversal of the smaller cone in the Bloch sphere in Fig. \ref{fig:bloch_sphere_illustration}(c) a large number of times $N$, seen as the bounces at the edge of the $c_g$-plot in Fig. \ref{fig:bloch_sphere_illustration}(e) (solid orange lines). During this operation, of duration $T = N 2\pi/\Omega_R\approx\frac{4\Delta \theta}{\Omega^2}$, the state acquires a phase $\theta = N\pi (1-\frac{\Delta}{\Omega_R}) \approx \frac{N\pi}{2}(\frac{\Omega}{\Delta})^2$, while the integrated excited state population is $\int_0^T |c_e(t)|^2 dt \approx \frac{2\theta}{\Delta}$. 

\subsubsection{Capture and release of the energy eigenstate \label{sec:cap_and_rel}}


As an alternative far-detuned interaction protocol, we start with a constant Rabi frequency, $\Omega$, and detuning, $\Delta$, but halfway through the first traversal of the cone trajectory in Fig. \ref{fig:bloch_sphere_illustration}(c) , we abruptly lower the detuning such that the state becomes parallel to the torque vector $\vec{W}$ and hence stays constant, see Fig. \ref{fig:bloch_sphere_illustration}(d), as it is an eigenstate of the Hamiltonian, which accumulates a dynamical phase proportional to the eigenenergy. In the $c_g$-plot, Fig. \ref{fig:bloch_sphere_illustration}(e) (dashed red line), this phase accumulation corresponds to a circular arc. At any time, reversal of the detuning to its initial value, $\Delta$, causes the state to resume the conical trajectory in Fig. \ref{fig:bloch_sphere_illustration}(c) and return to the ground state. In the limit when $\Delta \gg \Omega$, the duration of the operation is $T\approx \frac{2\Delta \theta}{\Omega^2}$, i.e., only half compared to the third operation, while the integrated excited state population is the same, $\int_0^T |c_e(t)|^2 dt \approx \frac{2\theta}{\Delta}$.

We note that the situation of far off-resonant driving is often analyzed by perturbation theory, which identifies the so-called AC Stark shift of the ground state energy level, proportional with $\Omega^2/\delta$. This quantity, indeed, characterizes the phase evolution in both schemes 3. and 4. But, in scheme 3, the phase accumulation is of a geometric origin, while in scheme 4, it is predominantly of a dynamical origin (see below). 



\subsection{The dynamic and geometric origin of the phase\label{sec:dyn_geo}}

Let us consider a time-dependent quantum state $\ket{\psi(t)}$. By introducing the phase argument $\theta(t) = \arg \braket{\psi(0)|\psi(t)}$ \cite{Samuel1988} and the so-called projected state $\ket{\lambda(t)} = e^{-i\theta(t)} \ket{\psi(t)}$ in the Schr\"odinger equation, we obtain \cite{Aharonov1987,Pati1995,Pati1995a}
\begin{align} \nonumber
\theta(t) 
&=
-\int_0^t \braket{\psi(t')|H(t')|\psi(t')} dt'\\
&+
i \int_0^t \braket{\lambda(t') | \dot{\lambda}(t')} dt',
\label{eq:phase_expression}
\end{align}
where $\ket{\dot{\lambda}} = \frac{d}{dt} \ket{\lambda}$. We refer to the first term on the right-hand side as the dynamic phase, $\theta_{\text{dyn}}$, and the second term as the geometric phase, $\theta_{\text{geo}}$. 

A two-level system that undergoes a cyclic evolution in time $T$, will accumulate a geometrical phase $\theta_{\text{geo}}(T) = \pm \frac{1}{2} A_S$, where $A_S$ denotes the area, or solid angle, enclosed on the Bloch sphere \cite{Berry1984,Aharonov1987} (see also Appendix~\ref{app:phases}), and the $\pm$ sign depends on the direction of the path. The total phase $\theta = \theta_{\text{dyn}}+\theta_{\text{geo}}$, however, includes also the value of $\theta_{\text{dyn}}$, which is easily of the same magnitude as $\theta_{\text{geo}}$. In fact, none of the phases are unique but depend on the chosen value of the ground state energy and the choice of interaction picture for the solution of Schr\"odinger´s equation. As discussed further in Sec. \ref{sec:interaction_picture} and Appendix~\ref{app:phases}, this ambiguity is purely formal and without consequences for predictions by the theory. But, it must necessarily be kept in mind in the design of processes that aim to control the phase on a quantum state. 

With the choice of interaction picture and ground state energy leading to Eq.~(\ref{eq:2level_Hamiltonian}), the processes with the Bloch vector trajectories depicted in Fig.~\ref{fig:bloch_sphere_illustration}(a, b, c), have vanishing dynamical phases. This occurs because the expectation value of the Hamiltonian, Eq.~(\ref{eq:2level_Hamiltonian}), vanishes for all times (the Hamiltonian is a conserved quantity, and its expectation value trivially vanishes in the initial ground state for all three processes). The phase accumulated on the ground state is hence given exclusively by the geometric areas, shown in the figures (the orange trajectory in panel (c) encircles the cone five times, accumulating a phase equal to five times the area shown). The process depicted in Fig.~\ref{fig:bloch_sphere_illustration}(d), however, yields both geometric, $\theta_{\text{geo}} = \pi (1-\frac{\Delta^c}{\Omega_R^c})$, and dynamic $\theta_{\text{dyn}} = (\Omega_R^l - \Delta^l) t^l / 2$ contributions to the phase, where $\Delta^c$ is the initial detuning used to traverse the cone (with generalized Rabi frequency $\Omega_R^c$), $\Delta^l$ is the detuning used when the Bloch vector is locked in space during a duration of $t^l$ (see Fig.~\ref{fig:bloch_sphere_illustration}(d)), and $\Delta^c = \Delta^l + \Omega_R^l$. Thus, despite the small red area being encircled only once by the Bloch vector, the process yields the same total phase change, $\theta = \pi/2$, as the processes depicted in the other panels. This phase accumulation is clearly visualized in the $c_g$-plot. We find it interesting to note that the familiar AC Stark shift mechanism applies a phase on the ground state with strikingly different, geometrical and dynamical, interpretations.

\subsection{The choice of interaction picture affects the values of the phases\label{sec:interaction_picture}}

As stated above, the phase acquired by the quantum state and the relative phase acquired by its different state components, depend on the (arbitrary) assignment of energy to the ground state and on the interaction picture chosen for the calculation. At this point, we recall that the two-level system addressed so far will be part of a larger system, and what matters in the end is not the phase of a single particular state, but the phase acquired by $\ket{g}$ compared to other ground states, or compared to the case where other processes may apply to the system. For instance, a simple change of phase of the laser drive may be incorporated in a corresponding interaction picture and formally yield a phase shift on the ground state. However, this does not necessarily imply the desired relative phase in the system.

\begin{figure}
    \includegraphics[width=\columnwidth]{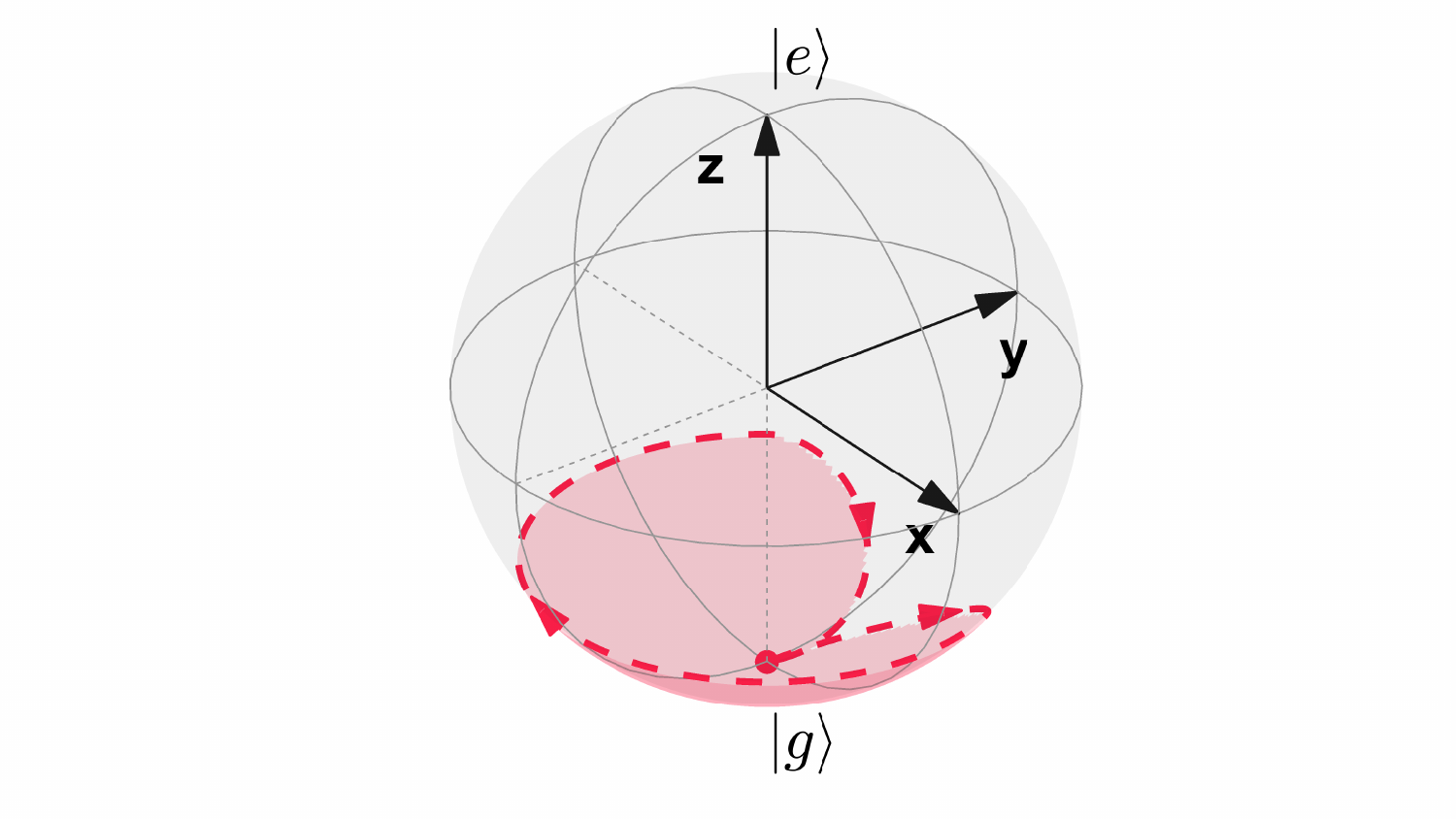}
    \caption{Depiction of the capture and release operation described in Sec. \ref{sec:cap_and_rel} in the frame that evolves with a fixed frequency equal to the atomic frequency, i.e., Eq.~(\ref{eq:2level_Hamiltonian}) applies with $\Delta = 0$ and the complex phase $\phi$ varies with time.}
    \label{fig:BS_reference_frame}
\end{figure}

The general ambiguity of the phase can be illustrated by picking a different convention and assigning energies $\pm \Delta/2$ to the excited and ground states (cf. the commonly applied two-level Hamiltonian $\Delta \hat{\sigma}_z/2$). Despite the fact that the physical dynamics remains unchanged, in this case, the ground state accumulates a phase even without an external drive. Of a somewhat more complex nature is the possibility to vary the detuning of the laser field with time, and hence obtain the phase of the laser field as the integral of the instantaneous frequency. We can apply the rotating wave approximation and choose a frame that evolves with the phase of the laser beam or a frame that evolves with a fixed frequency, e.g., equal to the atomic frequency. In the former case, Eq.~(\ref{eq:2level_Hamiltonian}) applies with fixed $\phi$ while the detuning $\Delta$ varies with time. In this frame, the operation shown in Fig. \ref{fig:bloch_sphere_illustration}(d) has $\theta_{\text{geo}} = \pi/10$ and $\theta_{\text{dyn}} = 4\pi/10$. In the latter case, $\Delta = 0$ and the complex phase $\phi$ varies with time, and the resulting trajectory on the Bloch sphere for the same operation is shown in Fig. \ref{fig:BS_reference_frame}, where $\theta_{\text{geo}} \approx 2\pi/6$ and $\theta_{\text{dyn}} \approx \pi/6$. In both frames the total phase is $\theta = \pi/2$. However, the choice of frame drastically alters the Bloch sphere trajectory and the magnitudes of the geometric and dynamic contributions. Since the different interaction pictures entail different phase factors on the excited basis state, they display different Bloch sphere trajectories, but the transformation between them does not affect the ground state amplitude, and the $c_g$-plot is identical in both frames (dashed red line in Fig. \ref{fig:bloch_sphere_illustration}(e)).

\subsection{Comparison of schemes}

\begin{figure}
    \includegraphics[width=\columnwidth]{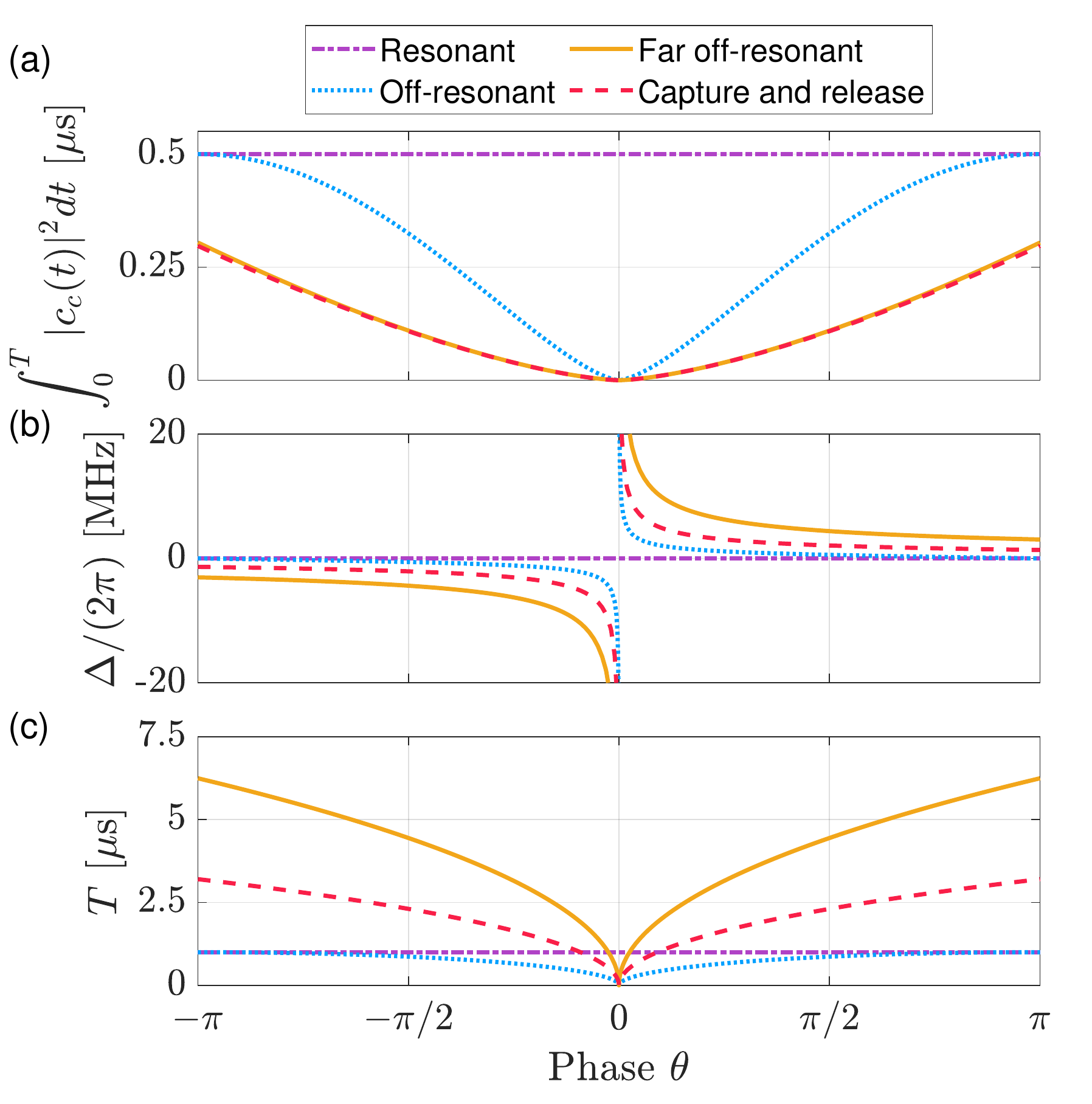}
    \caption{The (a) integrated excited state population, (b) detuning, and (c) total duration, of the four operations depicted in Fig. \ref{fig:bloch_sphere_illustration} are shown as a function of the phase $\theta$ added to $\ket{g}$. All operations use a constant Rabi frequency of $\Omega = \SI{1.0}{\mega \hertz}$, which determines all gate parameters for the resonant and off-resonant operations, while the far off-resonant operation performs $N=20$ loops. Finally, the capture and release gate uses the same initial detuning as the far off-resonant operation, and thus reach the same maximum excited state population. Note that we in (b) only depict the detuning used in the locked part of this gate operation.}
    \label{fig:path_comparison}
\end{figure}

Before moving on to more complex systems, we briefly compare the four strategies depicted in Fig.~\ref{fig:bloch_sphere_illustration} using a constant Rabi frequency $\Omega =\Omega_{\max} = \SI{1.0}{\mega \hertz}$. As already mentioned, the optimal strategy depends on which mechanisms and effects that one wants to eliminate or minimize. For this reason, we shall compare the four different strategies with respect to three generic properties. First, the integrated population of the excited state, assuming a high decay rate is a dominant source of error. Second, the detuning, which could drive unwanted transitions to spectator levels in the physical system. Third, the total gate duration, during which a control-qubit ion may decay and environmental interactions may cause dephasing and decoherence. We depict these three properties in Fig.~\ref{fig:path_comparison} (a), (b), and (c), respectively. The resonant drive has zero detuning and works relatively fast. However, it leads to a high population in the excited state. In contrast, the far off-resonant drive and the capture and release operations have small excited state populations, but require large detunings and longer gate times. Finally, the results of the off-resonant operation lies between these strategies for excited state population and detuning, and is generally faster. 

\section{\label{sec:lambda_tripod}Lambda and tripod systems}

In this section, we extend the analysis to systems with more levels. We first address how one can identify an effective two state system within a three-level lambda system. We can thus reuse the analyses and schemes of the previous section with the short-lived excited state replaced by a ground-excited superposition state. Any arbitrary unitary operation on the ground state qubit space can be implemented by putting a phase on a ground superposition state \cite{Roos2004}, and using a tripod system, this can also be accomplished as effective two-state dynamics between ground-ground and ground-excited superposition states. 

\subsection{\label{sec:lambda_phase}Putting a phase on the ground state in a lambda configuration}

\begin{figure}
    \includegraphics[width=\columnwidth]{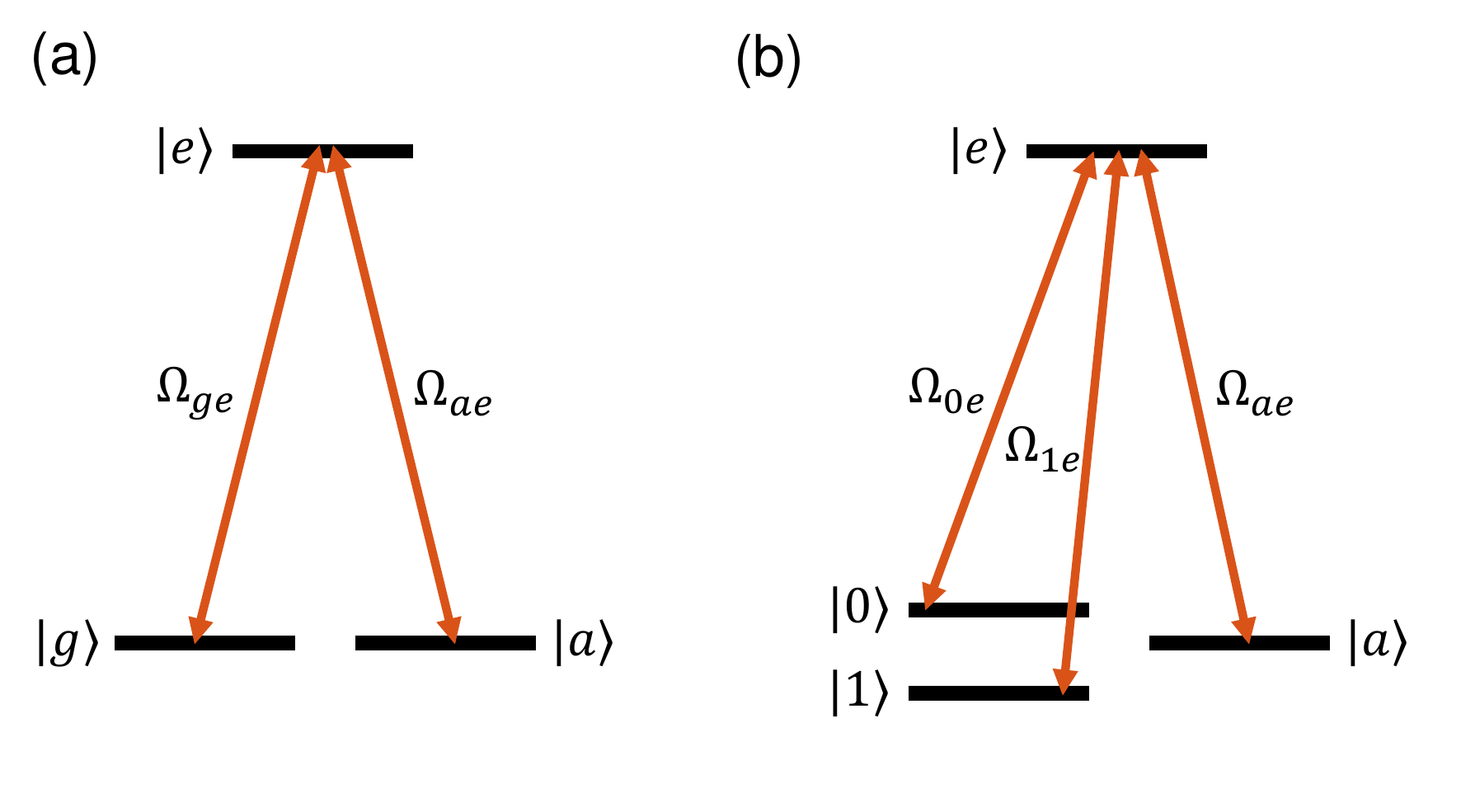}
    \caption{(a) Lambda three-level system with time-dependent complex Rabi frequencies $\Omega_{ge}(t)$ and $\Omega_{ae}(t)$. (b) Tripod four-level system with qubit levels $\ket{0}$ and $\ket{1}$, as well as a third ground state $\ket{a}$ and an excited state $\ket{e}$. }
    \label{fig:energy_levels}
\end{figure}

At the end of the previous section, we addressed the concern of using transitions to short-lived excited states for ground-state phase control. Direct coupling to another ground state would not suffer from excited state decay, but it would also not offer the frequency selective addressing of the optical transition, which is crucial for quantum computing schemes with rare-earth ions \cite{Kinos2021}. Hence, we are incentivized to find schemes that use the optical transitions to an excited state. In this section we extend the system by an ancillary ground state $\ket{a}$, which is only coupled via the excited state in a lambda configuration, see Fig.~\ref{fig:energy_levels}(a), such that we retain the qubit selective addressing. Following the discussion of Sec. \ref{sec:interaction_picture}, we can choose an interaction picture where the Hamiltonian has the form,
\begin{align}
    H^{\{e\,g\,a\}} =
    \frac{1}{2}
    \begin{bmatrix}
    0 & \Omega_{ge}(t)  & \Omega_{ae}(t) \\
    \Omega_{ge}^*(t)  & 0  & 0 \\
    \Omega_{ae}^*(t) & 0  & 0
    \end{bmatrix}
    \label{eq:H_ega}
\end{align}
and, possibly time-dependent, detunings are represented by the complex phases of the two Rabi frequencies $\Omega_{ge}(t)$ and $\Omega_{ae}(t)$. Before discussing phase operations, we note that assuming identical and constant Rabi frequencies implements an effective rotation of a spin-1 system in the $m=0,\pm 1$ basis, which may begin and end in the $m=1$ ground state. However, unlike the sign change by a complete rotation of a spin $1/2$ particle, the complete rotation of an integer spin yields no phase shift. 

Setting $\Omega_{ae}(t)=0$ trivially yields the two-level case already discussed in the previous section (now with $\Delta(t)$ incorporated in $\Omega_{ge}(t)$). However, the additional coupling enables a richer dynamics. For example, one can employ strong driving of the initially unpopulated states $\ket{e}$ and $\ket{a}$ to form dressed superposition states, $\alpha\ket{e} + \beta\ket{a}$ with different eigenenergies. Coupling the ground state $\ket{g}$ resonantly to the dressed state with a reduced excited state contents may thus yield a two-level dynamics that is less affected by excited state decay while yielding the same phase as the examples in the previous section. Additionally, we recall the STIRAP process \cite{Bergmann1998, Unanyan1999, Moller2007}, which employ destructive interference of the coherent couplings to avoid population of the intermediate unstable state. Deploying such destructive interference of the coherent excitation from the ground states $\ket{g}$ and $\ket{a}$ to the excited state $\ket{e}$ may yield a near-perfect ground state dynamics and phase change on $\ket{g}$.


Despite the differences between these methods and their origins, we can incorporate them into a single unified analysis. This analysis describes a broader class of unitary control strategies for the three-level lambda system that all return the system to its initial state $\ket{g}$ with a controllable phase $\theta$. As we will see, this analysis also allows us to apply elements of the two-level dynamics described in Sec. \ref{sec:phase}. 

\begin{figure*}
    \includegraphics[width=\textwidth]{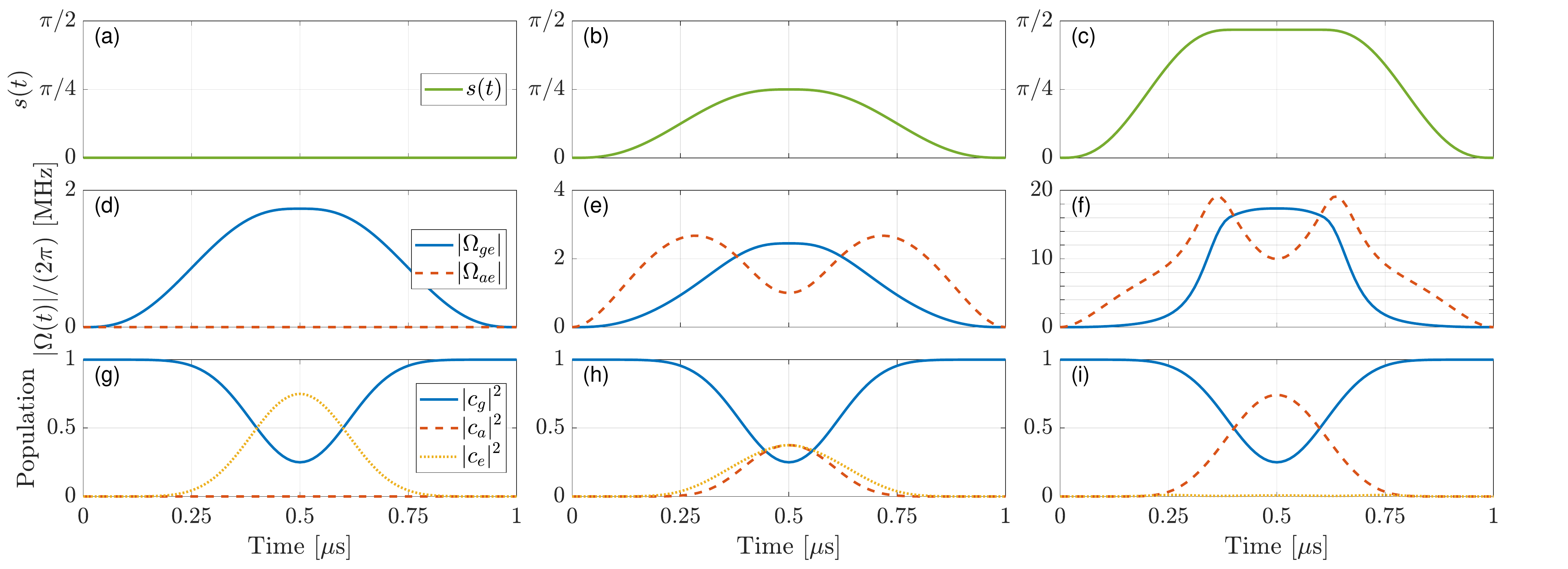}
    \caption{(a-c) Time-dependent shaping function $s(t)$, (d-f) Rabi frequencies, and (g-i) populations, for three different examples of possible lambda dynamics (columns) that return to state $\ket{g}$ with a phase $\theta=\pi/2$ upon completion. In all three examples, the dynamics in the two-level $\{\ket{g},\ket{p(t)}\}$-subspace is described by the off-resonant operation discussed in Sec. \ref{sec:off_res} and shown as the dotted blue lines in Fig. \ref{fig:bloch_sphere_illustration}(b, e), except state $\ket{e}$ should be replaced by $\ket{p(t)}$.}
    \label{fig:tripod_s_func}
\end{figure*}


We start by noting that any lambda system dynamics, e.g., described by Eq. (\ref{eq:H_ega}), can be expressed in a basis formed by the ground state $\ket{g}$ and two orthogonal superposition states, $\ket{p(t)}$ and $\ket{u(t)}$, of states $\ket{e}$ and $\ket{a}$, chosen such that $\ket{u(t)}$ is always unpopulated. This decomposition effectively reduces the system dynamics to two levels and allows us to reuse the strategies in Sec. \ref{sec:phase}. Based on the discussion in Sec. \ref{sec:interaction_picture}, we note that the superposition states $\ket{p(t)}$ and $\ket{u(t)}$ can, in a suitably defined frame of reference, be taken as real time-dependent superpositions of $\ket{e}$ and $\ket{a}$,
\begin{align}\label{eq:p_u_def}
    \ket{p(t)} &= \cos(s(t)) \ket{e} + \sin(s(t)) \ket{a}, \nonumber\\ 
    \ket{u(t)} &= -\sin(s(t)) \ket{e} + \cos(s(t)) \ket{a},
\end{align}
where $s(t)$ denotes a dimensionless, time-dependent shaping function that determines the complex Rabi drives $\Omega_{ge}(t)$ and $\Omega_{ae}(t)$, and vice versa, see Appendix \ref{app:tripod_derivation}. The populations of the bare states $\ket{e}$ and $\ket{a}$ are
\begin{align}\label{eq:c_e_pop}
    |c_e(t)|^2 &= \cos^2(s(t)) |c_p(t)|^2, \nonumber \\
    |c_a(t)|^2 &= \sin^2(s(t)) |c_p(t)|^2.
\end{align}

Fig. \ref{fig:tripod_s_func} shows three examples of the lambda system dynamics in analogy to the off-resonant operation described in Sec. \ref{sec:off_res} that puts a phase $\theta=\pi/2$ on the initial ground state $\ket{g}$. The panels in the left column depict the usual two-level dynamics ($s(t) = \Omega_{ae}(t)=0$), and thus only populates the $\{\ket{g}, \ket{e}\}$-subspace during the operation. The middle column describes an intermediate regime where $s(t)$ reaches $\pi/4$ in the middle of the operation, which implies that $\ket{p(T/2)}$ is an equal superposition of $\ket{e}$ and $\ket{a}$. In the right column, $s(t)$ is rapidly approaching $\pi/2$, which mostly confines the system in the $\{\ket{g}, \ket{a}\}$-subspace. The ideal shaping function $s(t)$ will depend on the relative role of different mechanisms in the experiment. For example, keeping $s(t)$ small permits use of weak Rabi frequencies at the cost of a high intermediate excited state population. In contrast, large values of $s(t)$ limit the excited state population but requires larger Rabi frequencies, similar to the requirements for STIRAP operations \cite{Bergmann1998, Unanyan1999}, including versions using shortcut to adiabaticity and superadiabatic transitionless driving \cite{Xi2010, Torrontegui2013, Baksic2016, Ribeiro2019}.

In all three cases, the path taken in the $c_g$-plot is identical to the dotted blue line in Fig. \ref{fig:bloch_sphere_illustration}(e), but the gradient color scale indicates the amplitude of being in the state $\ket{p(t)}$, and Eq. (\ref{eq:c_e_pop}) determines the populations in $\ket{e}$ and $\ket{a}$.



\subsection{From a phase to a gate in a lambda or tripod system}



We assume the simultaneous coherent driving of two qubit logic states $\ket{0}$ and $\ket{1}$, see Fig. \ref{fig:energy_levels}(b), as described by the Hamiltonian
\begin{align}
    H^{\{e\,0\,1\,a\}} =
    \frac{1}{2}
    \begin{bmatrix}
    0  & \Omega_{0e}(t) & \Omega_{1e}(t) & \Omega_{ae}(t) \\
    \Omega_{0e}^*(t) & 0 & 0 & 0\\
    \Omega_{1e}^*(t) & 0 & 0 & 0\\
    \Omega_{ae}^*(t) & 0 & 0 & 0\\
    \end{bmatrix}.
\end{align}

Let us first consider the lambda configuration where $\Omega_{ae}(t)=0$, and write the two other Rabi frequencies as
\begin{align}\label{eq:Omega_01}
    \Omega_{0e}(t) &= \Omega_{\tilde{1}e}(t) \sin(\eta/2), \nonumber\\
    \Omega_{1e}(t) &= \Omega_{\tilde{1}e}(t) \cos(\eta/2) e^{i\gamma}.
\end{align}
The two fields couple the (bright) coherent superposition state $\ket{\tilde{1}}=\sin(\eta/2) |0\rangle + \cos(\eta/2) e^{-i\gamma} |1\rangle$ to the excited state with Rabi frequency $\Omega_{\tilde{1}e}(t)$, while the (dark) superposition state $|\tilde{0}\rangle= \cos(\eta/2) |0\rangle - \sin(\eta/2) e^{-i\gamma} |1\rangle$ is uncoupled. $\ket{\tilde{1}}$ thus plays the same role as $\ket{g}$ in our two-state analysis, and driving the $\{\ket{\tilde{1}},\ket{e}\}$-transition may hence yield a complex phase $\theta$ on $\ket{\tilde{1}}$ relative to $\ket{\tilde{0}}$. By variation of the parameters $\theta,\eta$ and $\gamma$, this gives access to any unitary operation (up to a global phase factor) in the $\ket{0}$, $\ket{1}$ basis \cite{Roos2004}. 

Extending to the full tripod system shown in Fig. \ref{fig:energy_levels}(b) and applying three fields that effectively couple both $\ket{\tilde{1}}$ and $\ket{a}$ to the excited state $\ket{e}$, the operations discussed in Sec. \ref{sec:lambda_phase} are again sufficient to obtain a relative phase between $\ket{\tilde{1}}$ and $\ket{\tilde{0}}$, and thus perform a gate in the $\ket{0},\ket{1}$ basis.

\section{\label{sec:diss}Phases and mixed state dynamics}
Our discussion of the phase acquired by a certain basis state of a quantum system was carried out under the assumption of pure state dynamics. However, the role of dissipation is important and has to be assessed for applications, e.g., in quantum computing. Dissipation is often well described by a master equation which has to be solved for the density matrix of the system. For the two-level system the density matrix is in a one-to-one correspondence with a Bloch vector which explores also the interior of the Bloch spheres depicted in Fig. \ref{fig:bloch_sphere_illustration}(a-d). Unlike the pure state $c_g$ amplitude in Fig. \ref{fig:bloch_sphere_illustration}(e), the density matrix elements, and the Bloch vector components, do not change upon a global phase change. 

While it is, indeed, possible to assign geometric phases to mixed states \cite{Uhlmann1986, Sjoqvist2000, Tong2004}, quantum jump dynamics \cite{Carollo2003} and completely positive maps and quantum channels \cite{Ericsson2003, Kult2008}, a natural solution to properly assess the phase evolution is to return to its ultimate application, namely as a relative phase. In both pure and mixed state dynamics, what matters for physical observables is the coherence between a state and another state in the same system, or the same state in the case where the process does not occur (due to being controlled by another system). This implies an extension of the system by extra degrees of freedom and solving the master equation of the larger system with both Hamiltonian and damping terms. The dynamics of the global phase of $c_g$ is then formally replaced by the density matrix element $\rho_{g,f}$ where $\ket{f}$ is the reference state, potentially unaffected by the dynamics. 

If we only want to know how dissipation on our (target) system affects its phase dynamics, it is not necessary to solve the master equation for such a larger system. If the smaller target system solves a master equation of the Lindblad form
\begin{align}
\frac{d\rho}{dt}&=\frac{1}{i\hbar}[H,\rho] - \nonumber\\
    & \frac{1}{2}\sum_m (C_m^\dagger C_m \rho + \rho C_m^\dagger C_m) + \sum_m C_m \rho C_m^\dagger,
\end{align}
and we want to assess the phase evolution under one or another Hamiltonian $H_{A(B)}$ (for example driving or not driving a transition), this can be done by solving the two-sided master equation, 
\begin{align}
\frac{d\tilde{\rho}}{dt}&=\frac{1}{i\hbar}(H_A \tilde{\rho} - \tilde{\rho} H_B) - \nonumber\\
& \frac{1}{2}\sum_m (C_m^\dagger C_m \tilde{\rho} + \tilde{\rho} C_m^\dagger C_m) + \sum_m C_m \tilde{\rho} C_m^\dagger.
\end{align}
Assuming an initial matrix with $\tilde{\rho}_{g,g}(0)=1$ and all other elements vanishing, we may apply the same interpretation to $\tilde{\rho}_{g,g}(t)$ and plot it the same way as we did for $c_g(t)$ in the pure state case. We believe that this extension of the pure state analysis, and the corresponding trajectory of $\tilde{\rho}_{g,g}(t)$ in the complex plane, may offer useful insights in the interplay between dissipation and accumulation of phase.

\section{\label{sec:conc}Conclusion}

This article has discussed how to use coherent driving to effectively modify the phase of the complex ground state amplitude of a two-level atomic system. We showed by examples how the phase is composed of geometric and dynamic contributions. Their values depend on the interaction picture chosen for the calculation and any energy offset. The value of the total phase change can sometimes, but not generally, be deduced from an analysis of the Bloch vector dynamics. Hence, the assessment and potential optimization of the accumulated phase is not always intuitively simple. We found the plot of the relevant state amplitude $c_g(t)$ in the complex plane to be a useful supplement to the trajectory of the Bloch vector.

We then extended our discussion to the coherent dynamics of a three-level system, which inhabits richer dynamics. Still, our two-level analysis benefited the discussion of how to induce an arbitrary phase on a ground state amplitude in the larger system. Furthermore, we briefly discussed how general single-qubit gate operations can be accomplished by inducing a phase on a superposition state in three- and four-level systems. Controllable phase dynamics also finds applications in multi-qubit gates, which are often inspired and guided by intuition and experience from operations on simpler systems \cite{Isenhower2011,Khazali2020}, and which will only work if we have a proper understanding of the mechanisms behind the development of phases.

There are strong efforts with optimal control theory, shortcut to adiabaticity, and composite pulse methods to steer the state amplitudes and phases in different systems. We believe that our account of several aspects of the phase dynamics may benefit analyses and proposals employing such methods for quantum computing across all implementations of physical qubits, as well as contribute to the broader understanding of spin and phase dynamics, see also Ref. \cite{Wood2020}.

\begin{acknowledgments}
This research was supported by the Danish National Research Foundation (Grant agreement No. DNRF156), and has received funding from the European Union's Horizon 2020 research and innovation programme under grant agreement No. 820391 (SQUARE) and No. 754513 (Marie Sklodowska-Curie program).
\end{acknowledgments}

\appendix

\section{\label{app:harmonic_oscillator}The complex amplitude as an harmonic oscillator}
\begin{figure}
    \includegraphics[width=\columnwidth]{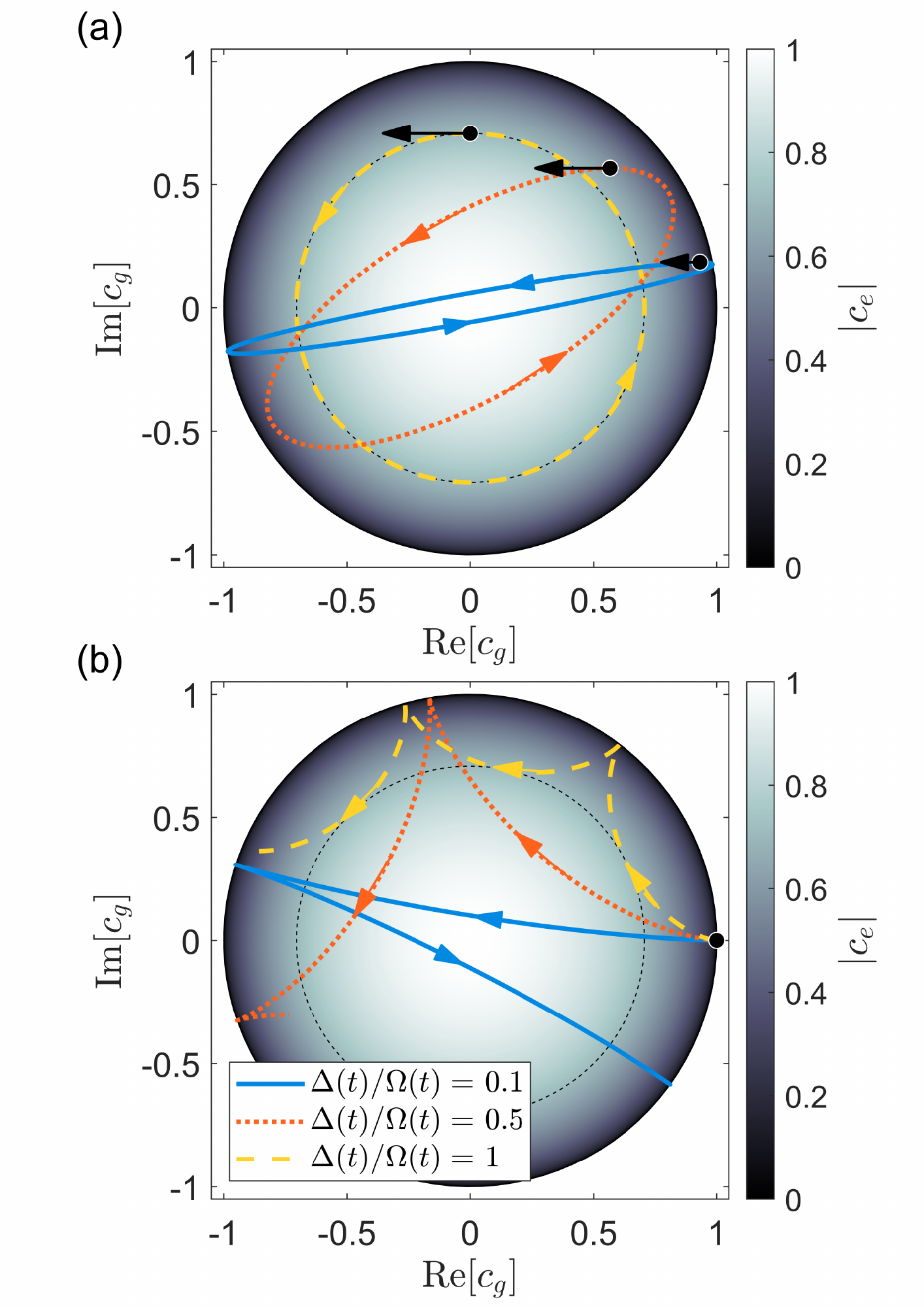}
    \caption{(a) Shows the trajectories for three different initial states, indicated by the black dots, when $\Delta(t) = 0$. The phase of $-i e^{-i\phi}c_e(0)$ is set so that the initial velocity, black arrows, points toward the left, and the speed is given by Eq. (\ref{eq:c_g_speed}). The trajectories are identical to how a long swinging pendulum would swing in the limit of small amplitudes. (b) Shows three examples with different ratios of $\Delta(t)/\Omega(t)$. The detuning acts as a steering force, turning the pendulum toward the right (left) if the detuning is positive (negative). Larger detunings results in a stronger steering force. }
    \label{fig:Pendulum_paths}
\end{figure}

Starting from a general two-level Hamiltonian on the form of Eq. (\ref{eq:2level_Hamiltonian}), but with a real time-dependent Rabi frequency and detuning, $\Omega = \Omega(t)$ and $\Delta = \Delta(t)$, the equations of evolution for the complex amplitudes are
\begin{align}\label{eq:rate_equations}
  \dot{c}_e(t) &= -i\frac{\Omega(t) e^{i\phi}}{2} c_g(t) - i \Delta(t) c_e(t),\nonumber\\
  \dot{c}_g(t) &= -i\frac{\Omega(t) e^{-i\phi}}{2} c_e(t).
\end{align}
The second derivative of $c_g(t)$ thus solves the equation
\begin{align}\label{eq:c_g_acceleration}
    \ddot{c}_g(t) = -\frac{\Omega(t)^2}{4} c_g(t) + \left(\frac{\dot{\Omega}(t)}{\Omega(t)} - i\Delta(t)\right) \dot{c}_g(t).
\end{align}
If we define $x(t) = \text{Re}[c_g]$ and $y(t)=\text{Im} [c_g]$, then the first term in this equation describes the restoring force of a 2-dimensional harmonic oscillator. Thus, if the Rabi frequency is constant and $\Delta(t) = 0$, the complex amplitude $c_g(t)$ will evolve like a long swinging pendulum in the limit of small amplitudes, see Fig. \ref{fig:Pendulum_paths}(a). The pendulum can at most swing a distance $1$ from the origin, which corresponds to the edge of the $c_g$-plot. Furthermore, the speed of the pendulum is always given by
\begin{align}\label{eq:c_g_speed}
    |\dot{c}_g(t)| = \frac{\Omega(t)}{2} |c_e(t)| = \frac{\Omega(t)}{2} \sqrt{1-|c_g(t)|^2}.
\end{align}
Thus, the speed is uniquely determined by the radial position of the pendulum, and the further out the pendulum is, the slower it moves. As given by Eq. (\ref{eq:rate_equations}), the phase of $-i e^{-i\phi}c_e(t)$ determines the phase of $\dot{c}_g(t)$, which in our analogy determines the direction of the pendulum's velocity. Thus, by choosing the phase of the driving field, $\phi$, one can choose the initial velocity direction of the pendulum, which determines its trajectory in the $c_g$-plot.

The second term in Eq. (\ref{eq:c_g_acceleration}) is a force acting in the direction of the velocity, and it can be shown that this does not alter the path of the pendulum, instead only affecting the speed at which the pendulum traverses the path, as given by Eq. (\ref{eq:c_g_speed}). Thus, ramping up or down the magnitude of the Rabi frequency does not affect the trajectory of the complex amplitude if $\Delta(t) = 0$. This is also true more generally, $\Delta(t)\neq0$, as long as the detuning is changed with the same factor as the Rabi frequency. 

Lastly, the third term in Eq. (\ref{eq:c_g_acceleration}) is a force acting perpendicular to the direction of the velocity. Thus, a detuning acts as a constant steering force pushing the pendulum to turn toward the right if $\Delta > 0$ or left if $\Delta < 0$, in relation to its current velocity direction, see Fig. \ref{fig:Pendulum_paths}(b).

\section{\label{app:phases} Discussion about phases}
As in the main text, we consider the phase accumulated by a state $\ket{\psi(t)}$ during its evolution, $\theta(t) = \arg \braket{\psi(0)|\psi(t)}$ \cite{Samuel1988}. Note that this definition, which is often attributed to Pancharatnam \cite{Pancharatnam1956}, is only well defined for non-orthogonal states $\braket{\psi(0)|\psi(t)} \neq 0$. Following Refs.~\cite{Aharonov1987,Pati1995,Pati1995a}, we may study the evolution of $\theta$ by introducing the projected state
\begin{align}
    \ket{\lambda(t)} = \frac{\braket{\psi(t)|\psi(0)}}{|\braket{\psi(t)|\psi(0)}|} \ket{\psi(t)}
    = e^{-i\theta(t)} \ket{\psi(t)}. 
\end{align}
Exploiting that $\ket{\psi}$ must obey the Schr\"odinger equation, we arrive at the equation
\begin{align}
    \dot{\theta}(t) = - \braket{\psi(t)|H(t)|\psi(t)} + i \braket{\lambda(t)|\dot{\lambda}(t)},
\end{align}
which has the solution $\theta(t) = \theta_{\text{dyn}}(t) + \theta_{\text{geo}}(t)$ given by Eq.~(\ref{eq:phase_expression}). If the projected state is expressed in a set of coordinates, $\boldsymbol{\xi}(t)$, we may rewrite the geometric phase as $\theta_{\text{geo}} = i \int_{\boldsymbol{c}} \braket{\lambda(\boldsymbol{\xi}) | \nabla_{\boldsymbol{\xi}} \lambda(\boldsymbol{\xi})} \cdot d\boldsymbol{\xi}$ \cite{Pati1995,Pati1995a}, where $\boldsymbol{c}$ denotes the path in parameter space. Written in this form, $\theta_{\text{geo}}$ can sometimes be given a geometric interpretation. Using Stoke’s theorem, Berry famously derived that a state of a two-level system that reverts to its original self ($\boldsymbol{\xi}(0) = \boldsymbol{\xi}(T)$) in time $T$, will accumulate a geometric phase corresponding to half the solid angle, or area $A_S$, it encloses on the Bloch sphere $\theta(T) = \pm \frac{1}{2}A_S$. Here $\pm$ depends on the direction of the path \cite{Berry1984,Aharonov1987}. Note that Berry's own considerations were limited to the adiabatic case. 

The accumulated phase of a state is not invariant under unitary transformations $\ket{\psi'} = R \ket{\psi}$ ($H' = RHR^{\dagger} + i \dot{R}R^{\dagger}$), since 
\begin{align}
\theta'(t) = \arg \braket{\psi(0)|R^{\dagger}(0) R(t)|\psi(t)} \neq \theta(t),
\end{align}
where the prime denotes values obtained for the transformed system. This inequality does not imply any discrepancy between the different pictures, as long as all observables have unchanged, consistent expectation values. Thus, global phases make no difference, and different relative phases should be absorbed by corresponding phases in matrix elements of the observable in the different pictures. 

While the time evolution operator $U(t)$ may be different in different pictures, the probability of finding the system in its initial state after a later time $t$ must be unchanged
\begin{align} \nonumber
|\braket{\psi(0)|\psi(t)}|^2 &= |\braket{\psi(0)|U(t)|\psi(0)}|^2\\ \nonumber
&= 
|\braket{\psi'(0)|U'(t)|\psi'(0)}|^2\\
& = 
|\braket{\psi(0)|R^{\dagger}(0)U'(t)R(0)|\psi(0)}|^2.
\end{align}
This must hold for all initial states and, hence, implies the relation
\begin{align}
U(t) = R^{\dagger}(0) U'(t) R(0) e^{-i\alpha(t)},
\label{eq:transformation_of_unitary}
\end{align}
where $\alpha(t)$ denotes a global phase factor. With this we have
\begin{align}
    \theta'(t) = \arg \braket{\psi'(0)|U'(t)|\psi'(0)} = \theta(t) + \alpha(t).
\end{align}
We can relate $\alpha(t)$ to the phase induced by a collective energy shift of the system $H' = H + E(t)\mathds{1}$, where $\mathds{1}$ denotes the identity operator. 

Using Eq.~(\ref{eq:phase_expression}) we may derive the new dynamic phase
\begin{align}
\theta_{\text{dyn}}'(t)
&= \theta_{\text{dyn}}(t) 
-i\int_0^t \braket{\psi(t')| R^{\dagger}(t') \dot{R}(t') |\psi(t')} dt',
\label{eq:dynamic_phase}
\end{align}
and using $\ket{\lambda'(t)} = e^{-i\alpha(t)} R(t) \ket{\lambda(t)}$ we may also derive the new geometric phase
\begin{align} \nonumber
\theta_{\text{geo}}'(t)
&= \theta_{\text{geo}}(t) +\alpha(t) \\
&+ i\int_0^t \braket{\psi(t')| R^{\dagger}(t') \dot{R}(t') |\psi(t')} dt'.
\label{eq:geometric_phase}
\end{align}
Eqs.~(\ref{eq:dynamic_phase}-\ref{eq:geometric_phase}) show that, except for the global phase $\alpha(t)$, the change in the dynamic phase corresponds to the exact opposite of the change in the geometric phase. Hence, what we perceive as the geometric (or dynamic) phase depends on the frame of reference. The fact that the geometric and dynamic phases are not invariant also challenges the view that quantum gates based on geometric phases are generally more robust.

Refs.~\cite{Kobe1990,Kobe1990a,Kendrick1992} constructed an expression for an invariant dynamic phase by distinguishing between the energy operator $\mathcal{E}$ that determines the energy spectrum $\mathcal{E} \ket{\psi_n} = E_n \ket{\psi_n}$ and the Hamiltonian that governs the system dynamics. The energy operator transforms as $\mathcal{E}' = R \mathcal{E} R^{\dagger}$ (without the $i\dot{R}R^{\dagger}$ part), which makes the energy spectrum invariant. The distinction between energy and Hamiltonian operators leads to an alternative definition of the dynamic phase $-\int_0^t \braket{\psi|\mathcal{E}|\psi} dt'$, and similarly an alternative definition of the geometric phase, which are both invariant under the transformation between interaction pictures. However, $\theta_{\text{geo}}$, using this definition, does not retain the geometric interpretation as half the enclosed area on the Bloch sphere (unless one sticks to the particular interaction picture where $H=\mathcal{E}$).

\section{\label{app:tripod_derivation}Derivation of Rabi frequencies in a lambda configuration that lead to effective two-state dynamics and puts a phase on the initial ground state} 

In this section we determine what time-dependent complex Rabi frequencies must be applied in order to induce a phase $\theta$ on the ground state $\ket{g}$ when using the interactions of the lambda system shown in Fig. \ref{fig:energy_levels}(a). The lambda system Hamiltonian can be rewritten in the $\{\ket{p(t)},\ket{g},\ket{u(t)}\}$-basis, where $\ket{p(t)}$ and $\ket{u(t)}$ are described in Eq. (\ref{eq:p_u_def}), and are by definition, respectively, populated and unpopulated during the operation. We write the Hamiltonian in this basis as
\begin{align}
    H^{\{p\,g\,u\}} = 
    \begin{bmatrix}
    \Delta(t) & \frac{\Omega(t)}{2} e^{i\phi} & \frac{\Omega_{up}(t)}{2} \\
    \frac{\Omega(t)}{2} e^{-i\phi} & 0  & \frac{\Omega_{ug}(t)}{2} \\
    \frac{\Omega_{up}^*(t)}{2} & \frac{\Omega_{ug}^*(t)}{2} & 0
    \end{bmatrix},
    \label{eq:H_pgu}
\end{align}
where $\Delta(t)$ and $\Omega(t)$ are real time-dependent functions that, e.g., vary as the examples discussed in Sec. \ref{sec:examples}. Since $\ket{u(t)}$ is by definition always unpopulated, the dynamics is completely described in the $\{\ket{p(t)},\ket{g}\}$-subspace and the following Schr\"odinger equation must hold
\begin{align}
    i \begin{bmatrix}
    \dot{c}_p(t) \\
    \dot{c}_g(t) \\
    0 \\
    \end{bmatrix}
    = 
    H^{\{p\,g\,u\}}
    \begin{bmatrix}
    c_p(t) \\
    c_g(t) \\
    0 \\
    \end{bmatrix}.
\end{align}
This results in a constraint on the Rabi frequencies coupling to $\ket{u(t)}$;
\begin{align}\label{eq:Omega_constraint1}
\Omega_{up}^*(t)c_p(t) + \Omega_{ug}^*(t)c_g(t) = 0.
\end{align}

Now a transformation can be performed which takes the Hamiltonian from the $\{\ket{p(t)},\ket{g},\ket{u(t)}\}$-basis to the $\{\ket{e},\ket{g},\ket{a}\}$-basis
\begin{align}
    V = 
    \begin{bmatrix}
    \cos(s(t)) & 0 & -\sin(s(t))\\
    0 & 1 & 0 \\
    \sin(s(t)) & 0 & \cos(s(t)) \\
    \end{bmatrix},
\end{align}
where $s(t)$ is the real time-dependent shaping function that determines the relative populations of $\ket{e}$ and $\ket{a}$ according to Eq. (\ref{eq:c_e_pop}). 

The Hamiltonian in Eq. (\ref{eq:H_pgu}) transforms according to
\begin{align}\label{eq:H_ega_trans}
    H^{\{e\,g\,a\}} = VH^{\{p\,g\,u\}}V^\dagger + i\dot{V}V^\dagger.
\end{align}
There should not be any direct coupling between $\ket{g}$ and $\ket{a}$ in $H^{\{e\,g\,a\}}$, and this results in another constraint on the Rabi frequency coupling $\ket{u(t)}$ with $\ket{g}$;
\begin{align}\label{eq:Omega_constraint2}
    \cos(s(t)) \Omega_{ug}(t) + \sin(s(t)) \Omega(t) e^{-i\phi} = 0.
\end{align}
If both constraints, Eqs. (\ref{eq:Omega_constraint1}) and (\ref{eq:Omega_constraint2}), are fulfilled, the transformed Hamiltonian of Eq. (\ref{eq:H_ega_trans}) determines what complex time-dependent Rabi frequencies $\Omega_{ge}(t)$ and $\Omega_{ae}(t)$ should be applied in order to perform the desired operation in the three-level lambda system shown in Fig. \ref{fig:energy_levels}(a) and described by Eq. (\ref{eq:H_ega}). Note that in general $H^{\{e\,g\,a\}}$ has non-zero energies of the $\ket{e}$ and $\ket{a}$ states, but these can always be incorporated into the complex Rabi frequencies $\Omega_{ge}(t)$ and $\Omega_{ae}(t)$. 

Since the general form of this Hamiltonian is rather complicated, we continue our investigation with the assumption that the ratio between the detuning $\Delta(t)$ and the Rabi frequency $\Omega(t)$ in the $\{\ket{p(t)},\ket{g}\}$-subspace is kept constant, i.e., $\Delta(t) = \alpha \Omega(t)$, and we also set $\phi = 0$. In order to fulfill the constraint of Eq. (\ref{eq:Omega_constraint1}), we must know $c_p(t)$ and $c_g(t)$, whose solutions are similar to those in Eq. (\ref{eq:bloch_dynamics}), except we now allow $\Omega(t)$ to be time-dependent;
\begin{align} 
    c_p(t) &= \frac{-i}{\beta}
	\sin \bigg(\frac{\beta \Lambda(t)}{2}\bigg) e^{-i\frac{\alpha \Lambda(t)}{2}}, \nonumber\\
	c_g(t) &= \Bigg( 
	\cos \bigg(\frac{\beta \Lambda(t)}{2}\bigg)
    + i \frac{\alpha}{\beta}
	\sin \bigg(\frac{\beta \Lambda(t)}{2}\bigg)
	\Bigg) e^{-i\frac{\alpha \Lambda(t)}{2} }, \nonumber\\
	\Lambda(t) &= \int_0^t \Omega(t') dt', 
	\label{eq:time_dependent_bloch_dynamics} 
\end{align}
where $\beta = \sqrt{1+\alpha^2}$. Under these assumptions the transformed Hamiltonian become
\begin{widetext}
\begin{align}
    H^{\{e\,g\,a\}} = 
    \begin{bmatrix}
    \alpha \Omega(t) & \frac{\Omega(t)}{2 \cos(s(t))} & -i\frac{ds}{dt} + \frac{\Omega(t)}{2}\tan(s(t)) \left(\alpha - \frac{i\beta}{\tan\left(\frac{\beta \Lambda(t)}{2}\right)} \right) \\
    \frac{\Omega(t)}{2 \cos(s(t))} & 0 & 0 \\
    i\frac{ds}{dt} + \frac{\Omega(t)}{2}\tan(s(t)) \left(\alpha + \frac{i\beta}{\tan\left(\frac{\beta \Lambda(t)}{2}\right)} \right) & 0 & 0 
    \end{bmatrix}.
    \label{eq:H_ega_full}
\end{align}
By comparing this Hamiltonian with Eq. (\ref{eq:H_ega}) one can identify the fields $\Omega_{ge}(t)$ and $\Omega_{ae}(t)$ that should be applied to perform the desired evolution in the $\{\ket{p(t)},\ket{g}\}$-subspace, namely
\begin{align}
    \Omega_{ge}(t) &= \frac{\Omega(t)}{\cos(s(t))} e^{i\alpha \Lambda(t)}, \nonumber\\
    \Omega_{ae}(t) &= \left(-2i\frac{ds}{dt} + \Omega(t)\tan(s(t)) \left(\alpha - \frac{i\beta}{\tan\left(\frac{\beta \Lambda(t)}{2}\right)} \right) \right)e^{i\alpha \Lambda(t)},
\end{align}
\end{widetext}
where the energy $\alpha \Omega(t)$ of the excited state in Eq. (\ref{eq:H_ega_full}) has been incorporated into the time-dependent complex Rabi frequencies via the factor $e^{i\alpha \Lambda(t)}$. These are the Rabi frequencies that are used in order to perform the operations discussed in Fig. \ref{fig:tripod_s_func} for three different choices of $s(t)$. 

For the operation to work perfectly, the constraints put forth in Eqs. (\ref{eq:Omega_constraint1}) and (\ref{eq:Omega_constraint2}) require that $s(t)$ is $0$ whenever $c_p(t) = 0$, i.e., at the start and end of the pulse $s = 0$ and therefore $\ket{p(t)}$ at those times coincides with the excited state. However, keeping $s(t)$ constant and capping the maximum Rabi frequencies also provides good solutions, even though the operation itself is no longer perfect. 

\bibliography{Ref_lib}

\end{document}